# Generalized two-temperature model for coupled phonon-magnon diffusion

Bolin Liao, Jiawei Zhou and Gang Chen[*]

Department of Mechanical Engineering, Massachusetts Institute of Technology, Cambridge, Massachusetts, 02139, USA

**Abstract**

We generalize the two-temperature model [Sanders and Walton, Phys. Rev. B, **15**, 1489 (1977)] for coupled phonon-magnon diffusion to include the effect of the concurrent magnetization flow. Working within the framework of Boltzmann transport equation, we derive the constitutive equations for coupled phonon-magnon transport driven by gradients of both temperature and external magnetic fields, and the corresponding conservation laws. Our equations reduce to the original Sanders-Walton two-temperature model under a uniform external field, but predict a new magnon cooling effect driven by a non-uniform magnetic field in a homogeneous single-domain ferromagnet. We estimate the magnitude of the cooling effect in yttrium iron garnet, and show it is within current experimental reach. With properly optimized materials, the predicted cooling effect can potentially supplement the conventional magnetocaloric effect in cryogenic applications in the future.

Spin caloritronics [1,2] is a nascent field of study that looks into the interaction between heat and spin. In addition to providing ways of thermally manipulating magnetization and magnetic domain walls [3–5] as supplements to conventional spintronics, it also holds promise of novel energy harvesting and cooling applications owing to the recent discovery of the spin Seebeck effect (SSE) [6–10] and its reciprocal spin Peltier effect (SPE) [11,12] . Despite existing debates on details, it has been widely recognized that the aforementioned spin caloritronic effects are consequences of the interactions between phonons, electrons and spins [13–16]. From this perspective, spin caloritronics is a natural extension of both thermoelectrics and spintronics. Phonons are responsible for heat conduction in most solids; in metallic and semiconducting materials, electrons are carriers of charge, heat and spin; in magnetic materials, magnons [17] – the collective excitations of spins – also participate in transporting spin [18] and heat [19]. Coupled to these carriers are thermodynamic forces that drive their flows [20]: the gradients of temperature, electrochemical potential and non-equilibrium magnetization [21]. For conditions close to equilibrium, it is particularly convenient to treat the coupled transport phenomena within the phenomenological framework of irreversible thermodynamics [20], where the Onsager reciprocity relation serves as the link between concurrent flows. Routinely used in studying the coupled transport of electrons and phonons [20–22], the method of irreversible thermodynamics has also been utilized in analyzing the coupled transport of heat and charge with spins [21,23–26] .

---

[*] To whom correspondence should be addressed: gchen2@mit.edu.

In this paper we limit our discussion to ferromagnetic insulators without free conducting electrons. Further steps to understanding the spin caloritronic effects require microscopic models that provide quantitative information of the transport processes, for example the kinetic coefficients [20] that connect the driving forces to the corresponding fluxes. For studying thermoelectrics, the coupled transport processes are typically treated within the framework of Boltzmann transport equation (BTE) [22], which in the diffusion regime gives quantitative kinetic coefficients, and is capable of delineating ballistic transport [27] when solved with proper boundary conditions. It is particularly a natural way to describe thermally induced transport processes where coherent contributions are not important. On the other hand, the spintronics community often uses the Landau-Lifshitz-Gilbert (LLG) [28] equation for the dynamics of magnetization. Compared with BTE, LLG adopts more a "wave-like" point of view, where the coherent dynamics is important and the thermal relaxation acts as a damping factor. Indeed the long wavelength magnons have been shown to exhibit macroscopic coherence lengths at room temperature [29], and LLG is necessary to account for their behaviors. For the thermal transport, however, magnons with a wide range of wavelengths and coherent lengths will be excited, and LLG seems no longer a particularly convenient description. A recent work by Hoffman *et al.* [30] applied a "semi-phenomenological" stochastic LLG equation to modeling the longitudinal SSE, where the temperature effect was incorporated via a thermally fluctuating Langevin field. Since a linear phonon temperature distribution was presumed in their work, it did not fully solve the coupled phonon-magnon transport problem. An alternative approach to this problem adopts a more "particle-like" picture. The pioneering work by Sanders and Walton [31] treated the coupled phonon-magnon thermal diffusion process with a two-temperature model, where phonons and magnons were modeled as two gases of bosons, each locally in thermal equilibrium with different temperatures, and the local energy exchange rate between them is proportional to the temperature difference. This model was later used to explain the spin Seebeck effect [13], and was recently extended to take into account the boundary heat and spin transfer [32]. It also served as a modeling tool for interpreting dynamic measurements of the thermal conductivity of spin ladder compounds [33,34] and the static direct measurement of the magnon temperature [35], and has been applied to other carrier systems such as electron-phonon [36] and acoustic-phonon-optical-phonon [37].

In their original formulation, Sanders and Walton did not consider the associated magnetization flow with the magnon heat flow. On the other hand, Meier and Loss [38] showed that the magnon flow could also be generated by a non-uniform external magnetic field, but they did not look into the thermal aspect of this transport process. In this paper we combine the two paths and give a unified description of the coupled phonon-magnon diffusive transport of both heat and magnetization, which is also applicable when the external magnetic field is non-uniform.

Magnons are (in most cases [34]) bosonic excitations, and in equilibrium obey the Bose-Einstein distribution:

$$f_0(\mathbf{r},\mathbf{k}) = \frac{1}{\exp\left[\dfrac{\hbar\omega(\mathbf{k}) + g\mu_B B(\mathbf{r})}{k_B T_m(\mathbf{r})}\right] - 1}, \qquad (1)$$

where $\hbar\omega(\mathbf{k})$ is the magnon energy [17,28] without external magnetic field, $g$ is the Landé g-factor, $\mu_B$ is the Bohr magneton ($-g\mu_B$ combined represents the amount of magnetic moment carried by a single magnon [38]), $T_m$ is the magnon temperature and $B$ is the external magnetic field. Here we neglect the magnetic dipolar interaction and magnetic anisotropy for simplicity. Although magnons can reach a quasi-equilibrium state with a finite chemical potential under parametric pumping [39], here we treat magnons with vanishing chemical potential for local equilibrium is assumed. Next we write down the steady-state Boltzmann transport equation with the relaxation time approximation (RTA):

$$-\frac{f-f_0}{\tau_m} = \mathbf{v} \cdot \nabla_{\mathbf{r}} f_0, \tag{2}$$

where $f(\mathbf{r},\mathbf{k})$ is the non-equilibrium distribution function of magnons, $\mathbf{v}(\mathbf{k})$ is the group velocity of magnons, $\tau_m = \left(\frac{1}{\tau_{m-m}} + \frac{1}{\tau_{m-p,ela}} + \frac{1}{\tau_{m-imp}}\right)^{-1}$ is a lumped relaxation time of magnons including effects of magnon-magnon scattering [40], elastic magnon-phonon scattering [41] and elastic magnon-impurity scattering [42]. The inelastic magnon-phonon scattering is responsible for the local energy exchange between magnons and phonons [41], and in general cannot be written in a relaxation-time form [22]. Thus we follow Sanders and Walton [31] here and consider the energy exchange process separately in the conservation laws later. We emphasize the validity of this separation requires that phonon-magnon interactions be much weaker than magnon-magnon interactions.

After obtaining the non-equilibrium distribution function $f(\mathbf{r},\mathbf{k})$, we can calculate the local magnetization and heat flows carried by magnons. The magnetization is $\mathbf{J}_m = -g\mu_B \int \frac{d^3\mathbf{k}}{(2\pi)^3} f\mathbf{v}$, where the minus sign accounts for the fact that the excitation of magnons reduces the total magnetization [43]: $\mathbf{M}(\mathbf{r}) = \mathbf{M}_s - g\mu_B n_m$, where $\mathbf{M}_s$ is the saturation magnetization, and $n_m$ is the number density of magnons. To calculate the magnon heat flow, we start with the thermodynamic relation of a magnet [20]: $dE = dQ + BdM = dQ - Bg\mu_B dn_m$, where $E$ is the total energy of the magnet *and* the interaction energy ($BM$) between the magnet and the magnetic field, and thus field-independent [44], corresponding to $\hbar\omega(\mathbf{k})$ microscopically (in contrast $\hbar\omega(\mathbf{k}) + g\mu_B B$ corresponds to the field-dependent "spectroscopic energy" [44]). Differentiating the above relation with respect to time, we get the magnon heat flux $\mathbf{J}_{qm} = \mathbf{J}_e - B\mathbf{J}_m = \int \frac{d^3\mathbf{k}}{(2\pi)^3}(\hbar\omega + g\mu_B B) f\mathbf{v}$, where $\mathbf{J}_{qm}$ is the magnon heat flux, $\mathbf{J}_e$ is the magnon energy flux. The term $B\mathbf{J}_m$ describes the transport of the magnetic interaction energy associated with the magnetization flow, analogous to $\varphi \mathbf{J}_c$ in the case of electrons, where $\varphi$ is the electro-static potential and $\mathbf{J}_c$ is the electrical charge flux. Combining the above expression with Eqs. (1) and (2), we arrive at the constitutive equations for the magnon transport:

$$-\mathbf{J}_m = L_{11}\nabla B + L_{12}(-\nabla T_m), \tag{3}$$

$$\mathbf{J}_{qm} = L_{12}T_m \nabla B + L_{22}(-\nabla T_m), \tag{4}$$

with the kinetic coefficients given by (assuming an isotropic magnon dispersion):

$$L_{11} = -\frac{(g\mu_B)^2}{3}\int_\omega \tau_m v^2 \frac{\partial f_0}{\partial(\hbar\omega)} D(\omega)d\omega, \tag{5}$$

$$L_{12} = -\frac{g\mu_B}{3T_m}\int_\omega (\hbar\omega + g\mu_B B)\tau_m v^2 \frac{\partial f_0}{\partial(\hbar\omega)} D(\omega)d\omega, \tag{6}$$

$$L_{22} = -\frac{1}{3T_m}\int_\omega (\hbar\omega + g\mu_B B)^2 \tau_m v^2 \frac{\partial f_0}{\partial(\hbar\omega)} D(\omega)d\omega, \tag{7}$$

where $D(\omega)$ is the magnon density of states. We can interpret $L_{11}$ as the isothermal magneto-conductivity $\sigma_m$ and $L_{22}$ as the uniform-field magnon thermal conductivity $\kappa_m$, and define $L_{12}$ as a magneto-thermal coupling coefficient $\varsigma_m$. Note that the Onsager reciprocity relation manifests itself explicitly in Eqs. (3) and (4). It can be shown using Cauchy-Schwartz inequality that $L_{11}L_{22} \geq T_m L_{12}^2$ (in the case of electron transport, this inequality implies a positive zero-current thermal conductivity [22]), which guarantees the net entropy generation in this system is non-negative [45]. Eqs. (3)-(7) are reminiscent of electron transport, and the external field $B$ seems to play a similar role as the electrochemical potential of electrons. We need to point out here, however, a critical difference between electrons and magnons. The number of electrons is conserved, thus the electrochemical potential includes the contribution of a finite chemical potential that can be "self-adjusted" during the transport process, whereas the number of magnons is not conserved, and the $B$ field does not have any internal contributions (given the magnetic dipolar interaction is negligible).

With the constitutive equations (3) and (4), we still need conservation laws to complete the formulation. We first look at the phonon system. At steady state, the phonon energy can either be transported by the phonon heat flux or transferred to the magnon system. Thus in the spirit of Sanders and Walton's original work, the phonon energy conservation states:

$$\nabla \cdot \mathbf{J}_{qp} = \frac{C_m C_p}{C_m + C_p} \frac{T_m - T_p}{\tau_{mp}} \equiv g_{mp}(T_m - T_p), \tag{8}$$

where $\mathbf{J}_{qp}$ is the phonon heat flux, $T_p$ is the phonon temperature, $C_m$ and $C_p$ are the volumetric specific heat of magnons and phonons, $\tau_{mp}$ is a phenomenological time scale characterizing the inelastic interaction between phonons and magnons, and we define $g_{mp}$ as a lumped coefficient of phonon-magnon energy exchange. It is worth mentioning that Eq. (8) is the result of inelastic phonon-magnon scattering and in principle can be derived from a full version of BTE, similar as in the case of electron-phonon coupling [22]. Another conservation law has to do with the energy input from an external power source.

The impression that a magnon flow can be generated by a non-uniform static magnetic field can be misleading because it violates the second law of thermodynamics: no energy is put into the system, while the magnon flow can potentially output work. In reality, when the magnetization of the magnet changes, an electromotive force (EMF) is induced in the electromagnet (e.g., a solenoid). To maintain the magnetic field, the current running through the electromagnet has to overcome this EMF and thus do work. It can be shown [44] that the work done by the current in this process is precisely equal to $BdM$. Hence the local creation and annihilation of magnons enables the energy exchange between the system (including the ferromagnet itself and its interaction with the field) and the external power supply. A local version of the above statement can be translated to $\nabla \cdot (\mathbf{J}_{qm} + B\mathbf{J}_m + \mathbf{J}_{qp}) = B\nabla \cdot \mathbf{J}_m$, or more explicitly:

$$\nabla \cdot \mathbf{J}_{qm} + \nabla B \cdot \mathbf{J}_m = g_{mp}(T_p - T_m). \qquad (9)$$

Now combining Eqs. (8), (9) with (3) and (4), and the normal Fourier law for phonon heat conduction: $\mathbf{J}_{qp} = -\kappa_p \nabla T_p$ ($\kappa_p$ is the phonon thermal conductivity), the governing equations for the temperature distributions of magnons and phonons read (considering 1-dimentional situations):

$$-\kappa_p \frac{\partial^2 T_p}{\partial x^2} = g_{mp}(T_m - T_p), \qquad (10)$$

$$-\kappa_m \frac{\partial^2 T_m}{\partial x^2} + \left(2\varsigma_m \frac{\partial B}{\partial x}\right)\frac{\partial T_m}{\partial x} + \varsigma_m \frac{\partial^2 B}{\partial x^2} T_m - \sigma_m \left(\frac{\partial B}{\partial x}\right)^2 = g_{mp}(T_p - T_m). \qquad (11)$$

Here we assume the applied temperature and magnetic field gradients are small and thus the transport coefficients are averaged values that do not explicitly depend on $T_m$ or $B$. Eqs. (10) and (11) reduce to the original Sanders-Walton model when the external magnetic field is uniform, even though in this case the magnetization flow is present ( $\mathbf{J}_m = -\varsigma_m \nabla T_m$ ).

More interesting phenomena emerge when non-uniform external magnetic field is applied. We expect a non-uniform external field will drive magnon flow, which is associated with a magnon heat flow, and cause temperature redistribution of both magnons and phonons due to the phonon-magnon coupling. Without a concise analytic solution with the coupling terms, we turn to numerical solutions for clarity, before which we first estimate the kinetic coefficients based on information in literature on yttrium iron garnet (YIG). Since the magnetic energy scale is pretty small ( $g\mu_B \approx 1.3$ K/T, for $g = 2$ in YIG), we expect the predicted effect to be more pronounced at low temperatures. Thus we use the low temperature expansion of the magnon dispersion $\hbar\omega(\mathbf{k}) = Dk^2a^2$, where $D \approx 1.8$ meV [46], and the lattice constant $a = 12.3$ Å for YIG [47]. For a similar reason, we neglect the field dependence of the kinetic coefficients in the following discussion. Further assuming a constant relaxation time $\tau_m$, we obtain the ratio $\frac{\sigma_m}{\varsigma_m} = \frac{g\mu_B}{k_B}\frac{\xi(1.5)}{\xi(2.5)} = 1.304$ K/T with $\xi(t) = \int_0^{+\infty} \frac{x^t e^x}{(e^x - 1)^2} dx$, which is analogous to the inverse of the Seebeck coefficient in the electron case. The value of $\tau_m$ is highly controversial [13], and here we adopt a value of $\tau_m \sim 1$ ns, which leads to the calculated uniform-field magnon thermal conductivity $\kappa_m \approx 8$ W/mK at 20K with zero field that is at least of the proper order of magnitude compared with the experiment [48]. With the same relaxation time, we obtain $\sigma_m \approx 0.25$ W/mT$^2$, and $\varsigma_m \approx 0.19$ W/mTK. For phonons, we choose $\kappa_p \approx 50$ W/mK [48]. At 20K, the specific heat of magnons and phonons are on the same order (~$10^4$ J/m$^3$K) [49]. Different claims on the value of $\tau_{mp}$ exists, ranging from below a few nanoseconds [32,50] to longer than a few hundred nanoseconds [13,51–53] at 300K. At lower temperature, this relaxation time will be longer, and we tentatively choose $\tau_{mp} \approx 100$ ns due to the large uncertainty of available data.

Provided the above parameters, we study numerically an experimentally realizable case: a strip of YIG (100 um long) connected to a thermal reservoir at 20K with one end, and the other end isolated. If part of the YIG strip is covered by a magnetic shielding material (such as µ-metal), a step-like magnetic field can be realized within YIG just by applying a uniform field. We model this step-like magnetic field as a smeared-out Fermi-Dirac function as shown in Fig. 1(a), and calculate the phonon temperature at the isolated end. In this case we apply adiabatic boundary conditions for magnons at both ends ( $\mathbf{J}_e = \mathbf{J}_{qm} + B\mathbf{J}_m = 0$ ). A phonon-temperature-drop of ~56 mK is predicted under a step field varying from 0.5T to 1.5T, with the temperature distribution of both phonons and magnons shown in Fig. 1(b). This temperature drop can be further amplified by increasing the field gradient as illustrated in Fig. 1(c). We would like to emphasize that the estimation here is very rough due to the lack of information, and is only intended to demonstrate a probable order of magnitude of this effect. The calculation above indicates that this magnon cooling effect may be detected under currently available experimental resolution. In passing we note that the predicted effect differs from the conventional magnetocaloric effect [54], such as adiabatic demagnetization, in that the magnetocaloric effect utilizes thermodynamic properties of the magnet (i.e. the field-dependent specific heat) in equilibrium, and a uniform field is often applied.

We provide another example where the magnon cooling effect is set up in close analogy to a thermoelectric Peltier cooling unit and calculate the coefficient of performance (COP) and effective zT. In this example the YIG strip is sandwiched between two thermal reservoirs with temperatures $T_h > T_c$, when a step field (as in Fig. 1(a)) is applied. The temperature profiles when $T_h = 20$ K, $T_h - T_c = 30$ mK and $(B_0, B_1) = (0.5\ \text{T}, 1.5\ \text{T})$ are plotted in Fig. 2(a), and it is clearly shown that heat is moved from the cold source to the hot source. The COP can be calculated via $\text{COP} \equiv \dfrac{Q_c}{W} = \dfrac{\mathbf{J}_{qp,cold}}{-\int_0^L B\nabla \cdot \mathbf{J}_m\, dx}$, and is plotted in Figs. 2(b) and 2(c) against varying temperature and field difference. At the fixed temperature difference of 30 mK (Fig. 2(b)), the optimal COP is around 2, corresponding to an equivalent thermoelectric module with ZT=0.01. From Fig. 2(c), the maximal attainable temperature difference is ~60 mK when $(B_0, B_1) = (0.5\ \text{T}, 1.5\ \text{T})$.

In summary, we have developed a semi-classical transport theory for coupled phonon-magnon diffusion. Our theory predicts that magnon flow can be driven by non-uniform magnetic field, and the heat carried by magnons associated with their flow can result in a cooling effect. In real materials, non-ideal effects such as magnetic dipolar interactions and the magnetic anisotropy need to be considered as a refinement to this work. We have estimated the magnitude of the magnon cooling effect in YIG, to show it can be verified by experiments. For practical uses, however, it is necessary to search for more suitable materials (preferably with lower thermal conductivities, and strong phonon-magnon interaction), and optimize the material properties via engineering efforts. We envision this new effect could supplement the conventional magnetocaloric effect in cryogenic applications in future.

**Ackowledgements**

We thank Sangyeop Lee and Mehmet Onbasli for helpful discussions. This article is based upon work supported partially by $S^3TEC$, an Energy Frontier Research Center funded by the U.S. Department of Energy, Office of Basic Energy Sciences, under Award No. DE-FG02-09ER46577 (for understanding the coupled phonon-magnon transport), and partially by the Air Force Office of Scientific Research Multidisciplinary Research Program of the University Research Initiative (AFOSR MURI) via Ohio State University (for studying the potential application of magnon cooling at cryogenic temperatures).

**References:**

[1] G. E. W. Bauer, E. Saitoh, and B. J. van Wees, Nat. Mater. **11**, 391 (2012).
[2] S. R. Boona, R. C. Myers, and J. P. Heremans, Energy Environ. Sci. **7**, 885 (2014).
[3] K.-R. Jeon, B.-C. Min, S.-Y. Park, K.-D. Lee, H.-S. Song, Y.-H. Park, Y.-H. Jo, and S.-C. Shin, Sci. Rep. **2**, (2012).
[4] M. Hatami, G. E. W. Bauer, Q. Zhang, and P. J. Kelly, Phys. Rev. Lett. **99**, 066603 (2007).
[5] W. Jiang, P. Upadhyaya, Y. Fan, J. Zhao, M. Wang, L.-T. Chang, M. Lang, K. L. Wong, M. Lewis, Y.-T. Lin, J. Tang, S. Cherepov, X. Zhou, Y. Tserkovnyak, R. N. Schwartz, and K. L. Wang, Phys. Rev. Lett. **110**, 177202 (2013).
[6] K. Uchida, S. Takahashi, K. Harii, J. Ieda, W. Koshibae, K. Ando, S. Maekawa, and E. Saitoh, Nature **455**, 778 (2008).
[7] C. M. Jaworski, J. Yang, S. Mack, D. D. Awschalom, J. P. Heremans, and R. C. Myers, Nat. Mater. **9**, 898 (2010).
[8] K. Uchida, J. Xiao, H. Adachi, J. Ohe, S. Takahashi, J. Ieda, T. Ota, Y. Kajiwara, H. Umezawa, H. Kawai, G. E. W. Bauer, S. Maekawa, and E. Saitoh, Nat. Mater. **9**, 894 (2010).
[9] K. Uchida, H. Adachi, T. An, T. Ota, M. Toda, B. Hillebrands, S. Maekawa, and E. Saitoh, Nat. Mater. **10**, 737 (2011).
[10] C. M. Jaworski, R. C. Myers, E. Johnston-Halperin, and J. P. Heremans, Nature **487**, 210 (2012).
[11] J. Flipse, F. L. Bakker, A. Slachter, F. K. Dejene, and B. J. van Wees, Nat. Nanotechnol. **7**, 166 (2012).
[12] J. Flipse, F. K. Dejene, D. Wagenaar, G. E. W. Bauer, J. B. Youssef, and B. J. van Wees, arXiv:1311.4772 (2013).
[13] J. Xiao, G. E. W. Bauer, K. Uchida, E. Saitoh, and S. Maekawa, Phys. Rev. B **81**, 214418 (2010).
[14] H. Adachi, J. Ohe, S. Takahashi, and S. Maekawa, Phys. Rev. B **83**, 094410 (2011).
[15] H. Adachi, K. Uchida, E. Saitoh, and S. Maekawa, Rep. Prog. Phys. **76**, 036501 (2013).
[16] K. S. Tikhonov, J. Sinova, and A. M. Finkel'stein, Nat. Commun. **4**, (2013).
[17] C. Kittel and C. Y. Fong, *Quantum Theory of Solids* (Wiley, New York, 1987).
[18] B. Wang, J. Wang, J. Wang, and D. Y. Xing, Phys. Rev. B **69**, 174403 (2004).
[19] C. Hess, Eur. Phys. J. Spec. Top. **151**, 73 (2007).
[20] H. B. Callen, *Thermodynamics and an Introduction to Thermostatistics* (Wiley, New York, 1985).
[21] M. Johnson and R. H. Silsbee, Phys. Rev. B **35**, 4959 (1987).
[22] G. Chen, *Nanoscale Energy Transport and Conversion: A Parallel Treatment of Electrons, Molecules, Phonons, and Photons* (Oxford University Press, Oxford; New York, 2005).
[23] W. M. Saslow, Phys. Rev. B **76**, 184434 (2007).
[24] W. M. Saslow and K. Rivkin, J. Magn. Magn. Mater. **320**, 2622 (2008).
[25] M. R. Sears and W. M. Saslow, Phys. Rev. B **85**, 035446 (2012).
[26] A. A. Kovalev and Y. Tserkovnyak, Europhys. Lett. **97**, 67002 (2012).
[27] G. Chen, Phys. Rev. B **57**, 14958 (1998).

[28] D. D. Stancil and A. Prabhakar, *Spin Waves Theory and Applications* (Springer, New York, 2009).
[29] Y. Kajiwara, K. Harii, S. Takahashi, J. Ohe, K. Uchida, M. Mizuguchi, H. Umezawa, H. Kawai, K. Ando, K. Takanashi, S. Maekawa, and E. Saitoh, Nature **464**, 262 (2010).
[30] S. Hoffman, K. Sato, and Y. Tserkovnyak, Phys. Rev. B **88**, 064408 (2013).
[31] D. J. Sanders and D. Walton, Phys. Rev. B **15**, 1489 (1977).
[32] M. Schreier, A. Kamra, M. Weiler, J. Xiao, G. E. W. Bauer, R. Gross, and S. T. B. Goennenwein, Phys. Rev. B **88**, 094410 (2013).
[33] M. Montagnese, M. Otter, X. Zotos, D. A. Fishman, N. Hlubek, O. Mityashkin, C. Hess, R. Saint-Martin, S. Singh, A. Revcolevschi, and P. H. M. van Loosdrecht, Phys. Rev. Lett. **110**, 147206 (2013).
[34] G. T. Hohensee, R. B. Wilson, J. P. Feser, and D. G. Cahill, Phys. Rev. B **89**, 024422 (2014).
[35] M. Agrawal, V. I. Vasyuchka, A. A. Serga, A. D. Karenowska, G. A. Melkov, and B. Hillebrands, Phys. Rev. Lett. **111**, 107204 (2013).
[36] G. Chen, J. Appl. Phys. **97**, 083707 (2005).
[37] R. B. Wilson, J. P. Feser, G. T. Hohensee, and D. G. Cahill, Phys. Rev. B **88**, 144305 (2013).
[38] F. Meier and D. Loss, Phys. Rev. Lett. **90**, 167204 (2003).
[39] S. O. Demokritov, V. E. Demidov, O. Dzyapko, G. A. Melkov, A. A. Serga, B. Hillebrands, and A. N. Slavin, Nature **443**, 430 (2006).
[40] F. J. Dyson, Phys. Rev. **102**, 1217 (1956).
[41] K. P. Sinha and U. N. Upadhyaya, Phys. Rev. **127**, 432 (1962).
[42] J. Callaway and R. Boyd, Phys. Rev. **134**, A1655 (1964).
[43] T. Holstein and H. Primakoff, Phys. Rev. **58**, 1098 (1940).
[44] C. Kittel, *Elementary Statistical Physics* (Wiley, New York, 1961).
[45] See online supplementary material for a brief discussion.
[46] A. B. Harris and H. Meyer, Phys. Rev. **127**, 101 (1962).
[47] R. L. Douglass, Phys Rev. **120**, 1612 (1960).
[48] R. L. Douglass, Phys. Rev. **129**, 1132 (1963).
[49] M. Guillot, F. Tchéou, A. Marchand, P. Feldmann, and R. Lagnier, Z. Für Phys. B Condens. Matter **44**, 53 (1981).
[50] N. Roschewsky, M. Schreier, A. Kamra, F. Schade, K. Ganzhorn, S. Meyer, H. Huebl, S. Geprägs, R. Gross, and S. T. B. Goennenwein, arXiv: 1309.3986 (2013).
[51] E. G. Spencer and R. C. LeCraw, Phys. Rev. Lett. **4**, 130 (1960).
[52] E. G. Spencer and R. C. LeCraw, Proc. IEE - Part B Electron. Commun. Eng. **109**, 66 (1962).
[53] C. Vittoria, P. Lubitz, P. Hansen, and W. Tolksdorf, J. Appl. Phys. **57**, 3699 (n.d.).
[54] A. M. Tishin and Y. I. Spichkin, *The Magnetocaloric Effect and Its Applications* (Institute of Physics Pub., Bristol; Philadelphia, 2003).

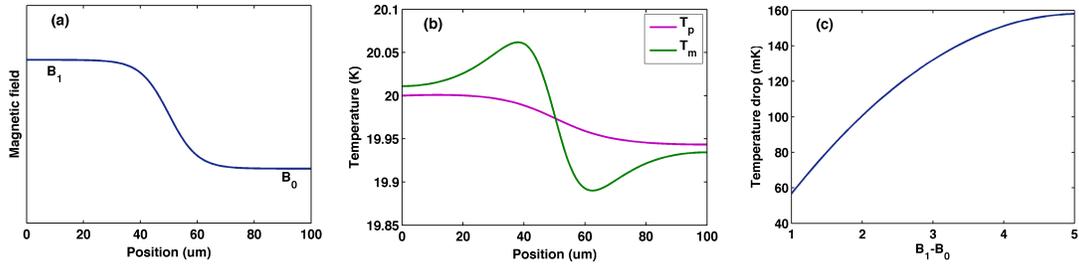

Figure 1. (a) The step-like magnetic field, smeared out as a Fermi-Dirac function. $B_0$ is fixed to be 0.5T in the following calculation. (b) The temperature distribution of phonons and magnons when $B_1 = 1.5$ T. One end of the sample ($x = 0$) is thermally connected to a reservoir at 20K, and the other end is isolated. (c) The dependence of the phonon temperature difference between the two ends on the difference of the magnetic field when $B_0$ is set to 0.5T.

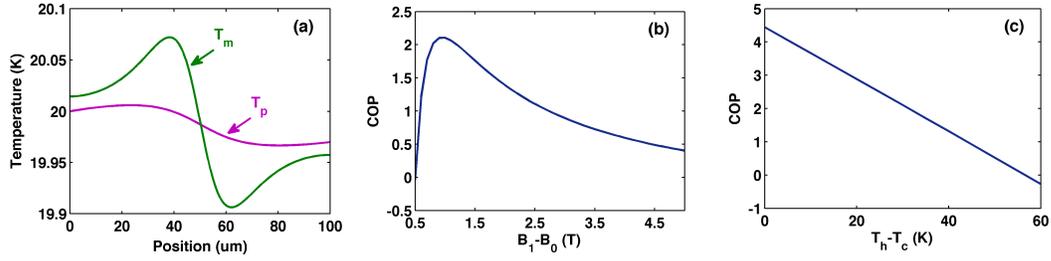

Figure 2. (a) The temperature profiles of phonons and magnons when $T_h = 20$ K, $T_h - T_c = 30$ mK and $(B_0, B_1) = (0.5 \text{ T}, 1.5 \text{ T})$. (b) COP versus the change of magnetic field when the temperature difference is fixed at 30mK. The hot-side temperature is fixed at 20K. (c) COP versus the temperature difference when the hot-side temperature is fixed at 20K and the magnetic field is fixed at $(B_0, B_1) = (0.5 \text{ T}, 1.5 \text{ T})$.

# Supplementary Material

## 1. Entropy generation

Carrier transport is a non-equilibrium process, thus usually irreversible and associated with entropy generation. Here we calculate the local entropy generation rate associated with the coupled phonon-magnon transport process, and prove it is always non-negative, thus obeying the second law of thermodynamics.

The entropy is transported with heat flow, and the divergence of entropy flow gives the local entropy generation rate at steady state [1]:

$$\dot{s} = \nabla \cdot \left(\frac{\mathbf{J}_{qm}}{T_m}\right) + \nabla \cdot \left(\frac{\mathbf{J}_{qp}}{T_p}\right)$$

$$= \frac{\sigma_m (\nabla B)^2}{T_m} + \kappa_m \left(\frac{\nabla T_m}{T_m}\right)^2 - \frac{2\zeta_m \nabla B \cdot \nabla T_m}{T_m} + \kappa_p \left(\frac{\nabla T_p}{T_p}\right)^2 + g_{mp}\left(\frac{T_m}{T_p} + \frac{T_p}{T_m} - 2\right). \quad (S1)$$

The first three terms represent the entropy generation due to friction and heat conduction in the magnon system, the 4$^{th}$ term represents the entropy generation due to heat conduction in the phonon system, and the last term represents the entropy generation caused by phonon-magnon energy exchange. Using the inequality $a + b \geq 2\sqrt{ab}$ ($a$ and $b$ are non-negative real numbers), we can show:

$$\frac{\sigma_m (\nabla B)^2}{T_m} + \kappa_m \left(\frac{\nabla T_m}{T_m}\right)^2 \geq 2\sqrt{\frac{\sigma_m \kappa_m}{T_m}} \frac{\nabla B \cdot \nabla T_m}{T_m} \geq \frac{2\zeta_m \nabla B \cdot \nabla T_m}{T_m}, \quad (S2)$$

$$\frac{T_m}{T_p} + \frac{T_p}{T_m} \geq 2. \quad (S3)$$

For (S2), we used the inequality $\kappa_m \sigma_m \geq T_m \zeta_m^2$ (or $L_{11}L_{22} \geq T_m L_{12}^2$). Thus the entropy generation rate in this system is always non-negative, an explicit embodiment of the second law of thermodynamics.